\documentclass{aa}

\usepackage{txfonts}
\usepackage{graphicx}
\usepackage{natbib}
\usepackage[usenames,dvipsnames]{color}
\bibpunct{(}{)}{;}{a}{}{,}

\begin{document}

\title{The VIMOS VLT Deep Survey\thanks{Based on data obtained
with the European Southern Observatory Very Large Telescope, Paranal,
Chile, under Large Programmes 070.A-9007 and 177.A-0837.  Based on
observations obtained with MegaPrime/MegaCam, a joint project of CFHT
and CEA/DAPNIA, at the Canada-France-Hawaii Telescope (CFHT) which is
operated by the National Research Council (NRC) of Canada, the
Institut National des Sciences de l'Univers of the Centre National de
la Recherche Scientifique (CNRS) of France, and the University of
Hawaii. This work is based in part on data products produced at
TERAPIX and the Canadian Astronomy Data Centre as part of the
Canada-France-Hawaii Telescope Legacy Survey, a collaborative project
of NRC and CNRS.}}
\subtitle{The different assembly history of passive and star-forming $L_B \gtrsim L_B^*$ galaxies in the group environment at $z < 1$}

\author{C. L\'opez-Sanjuan \inst{1,2,}\thanks{\email{clsj@cefca.es}}
\and O. Cucciati   \inst{1,3}
\and O.~Le F\`evre \inst{1}
\and L.~Tresse     \inst{1}
\and O.~Ilbert     \inst{1}
\and C.~Adami      \inst{1}
\and S.~Bardelli   \inst{3}
\and T.~Contini    \inst{4,5}
\and E.~Zucca      \inst{3}
}

\institute{Aix Marseille Universit\'e, CNRS, LAM (Laboratoire d'Astrophysique de Marseille) UMR 7326, 13388, Marseille, France 
\and Centro de Estudios de F\'{\i}sica del Cosmos de Arag\'on, Plaza San Juan 1, planta 2, 44001 Teruel, Spain 
\and INAF-Osservatorio Astronomico di Bologna, Via Ranzani 1, 40127, Bologna, Italy 
\and Institut de Recherche en Astrophysique et Plan\'etologie (IRAP), CNRS, 14 avenue \'Edouard Belin, 31400, Toulouse, France 
\and IRAP, Universit\'e de Toulouse, UPS-OMP, Toulouse, France 
}

\date{Submitted 05 March 2013 -- Accepted 15 August 2013} 

\abstract
{}{Understanding the role of environment in galaxy evolution is an important but still open issue. In the present work we study the close environment of red and blue $L_B \gtrsim L_B^*$ galaxies hosted by VVDS-Deep groups.}{We use the VIMOS VLT Deep Survey to study the close environment of galaxies in groups at $0.2 \leq z < 0.95$. Close neighbours of $L_B \gtrsim L_B^*$ galaxies ($M_B^{\rm e} = M_B + 1.1z \leq -20$) are identified with $M_B^{\rm e} \leq -18.25$ and within a relative distance $5h^{-1}$ kpc $\leq r_{\rm p} \leq 100h^{-1}$ kpc and relative velocity $\Delta v \leq 500$ km s$^{-1}$. The richness $\mathcal{N}$ of a group is defined as the number of $M_B^{\rm e} \leq -18.25$ galaxies belonging to that group. We split our principal sample into red, passive galaxies with $NUV - r \geq 4.25$ and blue, star-forming galaxies with $NUV - r < 4.25$. We study how the number of close neighbours per $L_B \gtrsim L_B^*$ galaxy depends on $\mathcal{N}$, colour, and redshift.}{Blue galaxies with a close companion are primarily located in poor groups, while the red ones are in rich groups. The number of close neighbours per red galaxy increases with $\mathcal{N}$, $\overline{n}_{\rm red} \propto 0.11\mathcal{N}$, while that of blue galaxies does not depend on $\mathcal{N}$ and is roughly constant. In addition, these trends are found to be independent of redshift, and only the average $\overline{n}_{\rm blue}$ evolves, decreasing with cosmic time.}{Our results support the following assembly history of $L_B \gtrsim L_B^*$ galaxies in the group environment: red, massive ($M_{\star} \sim 10^{10.8}\ M_{\odot}$) galaxies were formed in/accreted by the dark matter halo of the group at early times ($z \gtrsim 1$), therefore their number of neighbours provides a fossil record of the stellar mass assembly of groups, traced by their richness $\mathcal{N}$. On the other hand, blue, less massive ($M_{\star} \sim 10^{10.2}\ M_{\odot}$) galaxies have recently been accreted by the group potential and are still in their parent dark matter halo, having the same number of neighbours irrespective of $\mathcal{N}$. As time goes by, these blue galaxies settle in the group potential and turn red and/or fainter, thus becoming satellite galaxies in the group. With a toy quenching model, we estimate an infall rate of field galaxies into the group environment of $\Re_{\rm infall} = 0.9-1.5 \times 10^{-4}$ Mpc$^{-3}$ Gyr$^{-1}$ at $z \sim 0.7$.}


\titlerunning{The different assembly history of passive and star-forming $L_B \gtrsim L_B^*$ galaxies in the group environment}

\maketitle

\section{Introduction}\label{intro}
Since the discovery of the morphology -- density relation by \citet{Dressler80}, a number of investigations have aimed at a better understanding of the role of environment in the evolution of galaxies. Several works at $0.2 < z < 1.0$ studied how different galaxy properties depend on density and redshift. Some of these properties are the red, passive, and early-type fraction \citep[e.g.,][]{cucciati06,gerke07,cooper07,cucciati10den,tasca09,iovino10,vulcani11,grutz11}, the luminosity and mass function \citep[e.g.,][]{ilbert05,bolzonella10,pozzetti10}, or the merger fraction \citep{lin10,deravel09,deravel11,pawel12}. 
One of the most important results is that stellar mass seems to be driving most of the observed properties, most notably the higher fraction of red and early-type galaxies as mass increases. It is equally evident that environment is playing a significant role, particularly in participating in the transformation of blue galaxies with $M_{\star} \lesssim 10^{10.7}\ M_{\odot}$ into red ones \citep{cucciati10den,peng10}.

The density contrast, which is defined as the ratio of the local density excess around a galaxy with respect to the cosmological mean density at its redshift, is a continuous density tracer of the local environment from voids to the highest density peaks, which allows studying the dependence of galaxy properties on environment. However, the scale length on which environment affects galaxy properties the most is still a matter of debate.
Previous works have suggested that the observed effects of large-scale environment on galaxy properties are the mirror of the trends observed on much smaller scales \citep[e.g.,][and see also \citealt{wilman10}]{kauffmann04, blanton06, cucciati10den}. Nevertheless, it has not yet been verified that the large-scale environment does not play any role in shaping galaxy properties, and many times the explored small scales are not as small as desired, but simply the smallest allowed by the survey characteristics. Finding the scale on which environment affects the most galaxy properties would help, for instance, in disentangling the role of the different physical processes taking places on different scale lengths.

Another way to trace the effect of environment is to study the properties of galaxies residing in groups and clusters. Galaxy populations can be studied as a function of the distance from the group and cluster centre or of the group and cluster richness or mass, or can be compared with field galaxies. The advantage of this analysis is that groups and clusters can be directly related to the underlying dark matter (DM) structures, placing the analysis in a broader context. 

In the present paper we use close neighbours to provide an indication of the density field on small scales (less than 0.1$h^{-1}$ Mpc) of those galaxies hosted by groups of richness $\mathcal{N}$. This novel approach is physically motivated, since both close neighbours and richness are related to the underlying DM distribution. On the one hand, the number of close neighbours with a projected separation $r_{\rm p} \leq r_{\rm p}^{\rm max}$ in the sky plane, noted $\overline{n}$, is a proxy for the integral of the correlation function up to $r_{\rm p}^{\rm max}$ \citep{bell06,lin08,deravel09,robaina10}. The correlation function is a powerful tool for understanding the relation between galaxies and their host DM haloes. In this framework, halo occupation models have provided this connection in a phenomenological way \citep[e.g.,][]{cooray02,zheng05,abbas10,coupon12}. In these models, the so-called {\it one-halo component} governs the behaviour of the galaxy correlation function on scales smaller than $\lesssim 2 h^{-1}$ Mpc, while at larger separations galaxy correlations are dominated by the gravitational clustering of virialised DM haloes (the {\it two-halo component}). As a result, the number of close neighbours encodes information about the mass of the DM halo in which galaxies reside \citep[e.g.,][]{more11}. On the other hand, group richness $\mathcal{N}$ is tightly correlated with the DM halo mass of the groups (dispersion of the relation $\sigma \sim 0.2-0.3$ dex, \citealt{rozo09sig,rozo09N,andreon10}). A positive correlation between $\overline{n}$ and $\mathcal{N}$ for a given population therefore implies that this population is located in the DM halo of the group, while a lack of correlation suggests that the population is hosted by sub-haloes inside the more massive halo of the group. In the present paper we study the $\overline{n} - \mathcal{N}$ relation for red and blue $L_B \gtrsim L_B^*$ galaxies at $0.2 \leq z < 0.95$ by using the close neighbour candidates from \citet{clsj11mmvvds} and the group catalogue from \citet{cucciati10gr}, both obtained in the deep part of the VIMOS-VLT Deep Survey\footnote{http://www.oamp.fr/virmos/vvds.htm} (VVDS, \citealt{lefevre05}; \citealt{vvdsud}).

This paper is organised as follows. In Sect.~\ref{data} we summarise the VVDS-Deep data set, and in Sect.~\ref{richness} we describe the methodology for obtaining the richness of groups and the list of group members. In Sect.~\ref{nvsN} we estimate the number of close neighbours as a function of group richness, galaxy colour, and redshift, and we discuss the implications of our results in Sect.~\ref{discussion}. Finally, we present our conclusions in Sect.~\ref{conclusion}. We use $H_0 = 100h\ {\rm km\ s^{-1}\ Mpc^{-1}}$, $h = 0.7$, $\Omega_{m} = 0.3$, and $\Omega_{\Lambda} = 0.7$ throughout. All magnitudes refer to the AB photometric system \citep{oke83}. We assumed a \citet{chabrier03} initial mass function (IMF).

\section{The VVDS-Deep data set}\label{data}
The VVDS-Deep sample \citep{lefevre05,vvdsud} is magnitude-selected with $17.5 \leq I_{AB} \leq 24$. This spectroscopic survey has been conducted on the 0224-04 field with the VIMOS multi-slit spectrograph on the VLT \citep{lefevre03}, with 4.5h integration using the LRRED grism at a spectral resolution $R\sim230$. The multi-slit data processing has been performed using the VIPGI package \citep{scodeggio05}. Redshift measurement has followed a strict approach, with initial guesses based on cross-correlation with reference templates at the same redshift, followed by careful, independent eye-checking by two team members before comparing their results. The final redshifts and quality flags follow a statistically well-defined behaviour, leading to a survey for which at least 80\% of the sample has a secure spectroscopic redshift ($z_{\rm spec}$). This comprises sources with quality flag = 4 (100\% secure), 3 (98\% secure), 2 (84\% secure), and 9 (those with only a single secure spectral feature in emission in their spectrum, 72\% secure). The accuracy in the redshift measurement is 202 km s$^{-1}$ \citep{vvdsud}.

\begin{figure}[t!]
\resizebox{\hsize}{!}{\includegraphics{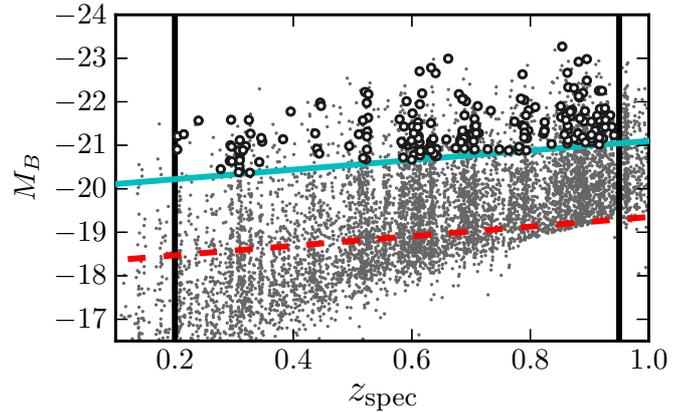}}
\caption{Rest-frame $B-$band absolute magnitude as a function of $z_{\rm spec}$ (dots) in the VVDS-Deep. Solid and dashed lines mark the principal ($M_B^{\rm e} = M_B + 1.1z \leq -20$) and companion ($M_B^{\rm e} \leq -18.25$) sample selections, respectively. Open circles are those principal galaxies located in groups. Vertical solid lines mark the redshift limits of the present study, $0.2 \leq z_{\rm spec} < 0.95$. [{\it A colour version of this plot is available in the electronic edition}].}
\label{mbvsz}
\end{figure}

Deep photometry is available in this field from a first campaign with the CFH12K camera in $BVRI$ (\citealt{lefevre04img} and \citealt{mccracken03}), followed by very deep observations with the CFHTLS survey in the $u^{*}g'r'i'z'$ bands \citep{cfhtlsT06} and with the WIRDS survey in $JHK_{s}$ bands \citep{WIRDS}. Using photometric redshifts, computed as in \citet{ilbert06phot}, we show that for the galaxies making up the 20\% incompleteness, about half have a tentative (quality flag = 1) spectroscopic redshift that is 48\% secure, and the other half have unknown spectroscopic redshifts (quality flag = 0), but we use photometric redshift estimates to fully understand the survey completeness as a function of magnitude and redshift. 

Several first-epoch VVDS-Deep galaxies with flag = 1 and 2 have been re-observed in the VVDS-Ultradeep ($I_{\rm AB} \leq 24.75$, \citealt{vvdsud}), providing a robust measurement of their redshift. This offers the opportunity to correct the redshift distribution of VVDS-Deep flag = 1 and 2 sources, bringing the final VVDS-Deep data set to the equivalent of 96\% completeness.

A total of 6445 galaxies with $0.2 \leq z_{\rm spec} < 0.95$ and $17.5 \leq I_{\rm AB} \leq 24$ 
(primary objects with flags = 1, 2, 3, 4, 9; and secondary objects, those that lie by chance in the slits, with flags = 21, 22, 23, 24, 29) from first- and second-epoch VVDS-Deep data \citep{vvdsud} have been used in this paper (Fig.~\ref{mbvsz}). We used flag = 1 sources, which are 48\% secure, when we search for kinematical close neighbours (Sect.~\ref{closepairs}), thanks to the improved weighting scheme in VVDS-Deep (see Sect.~\ref{complet}, and also \citealt{cucciati12} and \citealt{vvdsud} for more details). On the other hand, we used flag $\geq 2$ sources to identify galaxy groups (Sect.~\ref{groups}).

We derived the intrinsic luminosities and star formation rates ($SFR$) as fully described in \citet{cucciati12}. In summary, we used the full photometric information that is available to perform a fit of the spectral energy distribution (SED) with an updated version of the code {\it Le Phare\footnote{http://www.cfht.hawaii.edu/$\sim$arnouts/LEPHARE/lephare.html}}. The adopted method minimises the dependency on the template chosen for the SED fitting. The template library is from \citet{BC03}. Dust extinction was applied to the templates using the \citet{calzetti00} law. We derived stellar masses from the same SED fitting: the best fit template is normalised at one solar mass, and the stellar mass is the factor needed to rescale the template to the intrinsic luminosities.

\subsection{Kinematical close neighbours in the VVDS-Deep}\label{closepairs}
The distance between two sources can be measured as a function of their projected separation in the sky plane, $r_{\rm p}$, and of their rest-frame relative velocity along the line of sight, $\Delta v = {c|z_j - z_i|}/(1+z_i)$, where $z_i$ and $z_j$ are the redshift of the principal galaxy (the brightest galaxy in the pair) and of the companion galaxy, respectively. Two galaxies are defined as a close pair if $r_{\rm p}^{\rm min} \leq r_{\rm p} \leq r_{\rm p}^{\rm max}$ and $\Delta v \leq \Delta v^{\rm max}$. In our case we use $r_{\rm p}^{\rm min} = 5h^{-1}$ kpc, $r_{\rm p}^{\rm max} = 100h^{-1}$ kpc, and $\Delta v^{\rm max} = 500$ km s$^{-1}$. The lower limit in $r_{\rm p}$ is imposed to avoid spatial resolution limitations due to the size of the observed point spread function. The previous definition of ``close pair'' is widely used in merger fraction studies \citep[e.g.,][]{patton02,lin04,lin08,depropris07,deravel09,deravel11,clsj10pargoods} since 50\% to 70\% of the selected close pairs will finally merge \citep{patton00,patton08,bell06}. The close pair condition above selects close bound systems even when they are located in dense environments, such as the groups in the present work, but in these environments the probability of finding unbound close pairs increases \citep[e.g.,][]{lin10,jian12,pawel12}, providing information about the DM halo of the groups. Thus, in the present work we are interested in neither the merger fraction nor the merger rate, but instead in the total number of close neighbours around bright galaxies even if they will not merge with their principal galaxy. To avoid any confusion with merger studies, in the following we speak about {\it close neighbours} instead of {\it close pairs}.

\begin{figure}[t!]
\resizebox{\hsize}{!}{\includegraphics{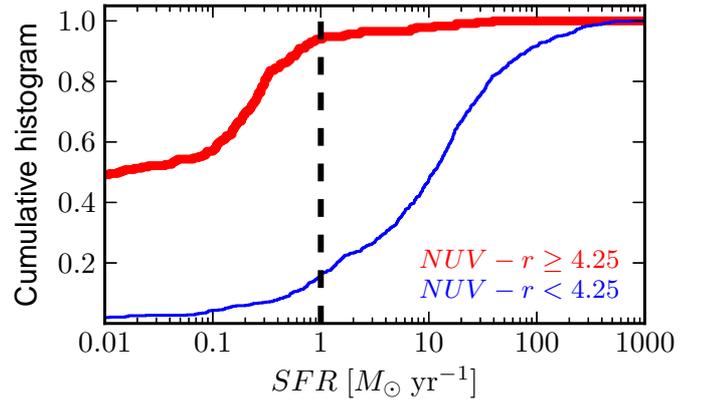}}
\caption{Cumulative histogram of red (thick line) and blue (thin line) principal galaxies as a function of their star formation rate. The definition of red and blue galaxy is labelled in the figure. The vertical dashed line marks $SFR = 1\ M_{\odot}$ yr$^{-1}$. [{\it A colour version of this plot is available in the electronic edition}].}
\label{histsfr}
\end{figure}

We select principal galaxies as those with an evolving rest-frame $B$-band absolute magnitude $M_B^{\rm e} \leq -20$, where $M_B^{\rm e} = M_B + Qz$ and the constant $Q = 1.1$ takes the evolution of the luminosity function in the VVDS-Deep into account \citep{ilbert05}. These principal galaxies are more luminous than $L_{B}^{*}$ at each redshift, since $M_B^{\rm e, *} \sim -20$ \citep{ilbert05}. To define close neighbours, we looked for those galaxies in the named companion sample that fulfils the close neighbour criterion for each galaxy of the principal sample. If one principal galaxy has more than one close neighbour, we took each possible pair separately (i.e., if the companion galaxies B and C are close to the principal galaxy A, we study the pairs A-B and A-C as independent). We also imposed a difference in the rest-frame $B$-band luminosity between the pair members of $L_{B,2}/L_{B,1} \geq 1/5$, which is an absolute $B$-band difference of $M_{B,2} - M_{B,1} \leq 1.75$. The companion sample therefore comprises galaxies with $M_B^{\rm e} \leq -18.25$. This ensures that our samples are complete in the redshift range under study, $0.2 \leq z < 0.95$ (Fig.~\ref{mbvsz}). There are 895 galaxies in the principal sample satisfying the $M_B^{\rm e} \leq -20$ selection, and 4156 galaxies with $M_B^{\rm e} \leq -18.25$ in which the neighbours of the principal galaxies have been searched for. In addition, in the following we refer to blue and red galaxies as those with rest-frame colour $NUV-r < 4.25$ and $NUV-r \geq 4.25$, respectively (see \citealt{arnouts07} and \citealt{clsj11mmvvds} for details about this selection). This colour is an excellent indicator of the present over past star formation activity. In Fig.~\ref{histsfr} we show the cumulative histogram of both red and blue principal galaxies as a function of their $SFR$. On the one hand, 95\% of the red principals have $SFR \leq 1\ M_{\odot}\ {\rm yr}^{-1}$ and 92\% have a specific star formation rate $sSFR = SFR/M_{\star} \leq 10^{-11}\ {\rm yr}^{-1}$. On the other hand, 16\% of the blue principals have $SFR \leq 1\ M_{\odot}\ {\rm yr}^{-1}$ and 6\% have $sSFR \leq 10^{-11}\ {\rm yr}^{-1}$. With this $sSFR$ limit, we can assume that red galaxies are {\it passive} and blue galaxies are {\it star-forming} \citep{schi07}.

Following the previous definitions, we find 64 close neighbours in the $\sim$0.5 deg$^2$ VVDS-Deep area\footnote{An extra red neighbour+neighbour pair is not included because we are interested in principal+neighbour statistics. That is, we find a triplet of principal red galaxies and discard the pair between the two fainter ones. We checked that our results are the same if this neighbour+neighbour pair is included.}, of which 24 have a red principal galaxy and 40 have a blue principal galaxy. We did not impose any colour selection to the neighbours.

\subsection{The VVDS-Deep group catalogue}\label{groups}
All the details about the methodology used to define reliable groups from spectroscopic surveys are in \citet{cucciati10gr}, so we only recall the main steps here.

The adopted group-finding algorithm is based on the Voronoi-Delaunay method \citep[][]{marinoni02}. In brief, a polyhedron is assigned to each galaxy, with a volume inversely proportional to the 3D density of galaxies in that given region (the higher the density, the smaller the polyhedron volume). Then, a three-phase group detection begins. In Phase I, group seeds are detected, when we searching for galaxies in a cylindrical window centred on the galaxies with the smallest Voronoi volumes. During Phase II, a larger cylindrical window is used to add members to the seeds, with the aim of recovering the central richness of the groups. Finally, Phase III consists in assigning the total number of members to the groups, when searching for galaxies residing in a cylindrical window with dimensions that scale with the central richness found in Phase II. 

\begin{figure}[t!]
\resizebox{\hsize}{!}{\includegraphics{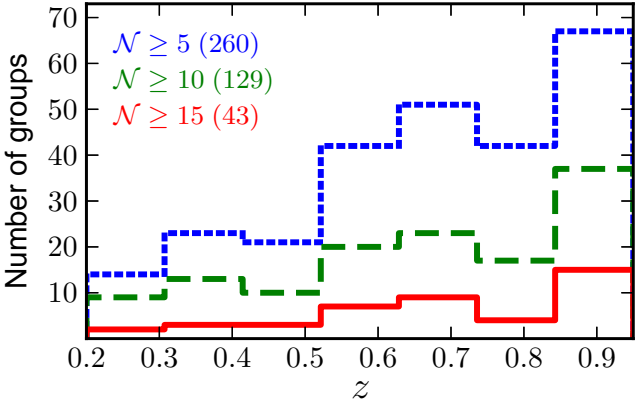}}
\caption{Redshift distribution of the VVDS-Deep groups with richness $\mathcal{N} \geq 5$ (260 groups, dotted line), $\mathcal{N} \geq 10$ (129 groups, dashed line) and $\mathcal{N} \geq 15$ (43 groups, solid line). $\mathcal{N}$ is defined as the number of weighted companion galaxies with $M_B^{\rm e} \leq -18.25$ that belong to the group (see text for details). [{\it A colour version of this plot is available in the electronic edition}].}
\label{histNfig}
\end{figure}

The algorithm has been fine-tuned using simulated mock galaxy catalogues taken from the MILLENNIUM Simulation \citep{springel05}, where the semi-analytical prescription of \cite{delucia07} were applied. These mock catalogues were prepared to reproduce the same VVDS observational strategy (flux limits, sampling rate, field geometry, etc.). The fine-tuning had the aim of maximising the completeness and the purity of the resulting group catalogue, and to minimise possible selection effects. In particular, the number density distribution as a function of both redshift, $n(z)$, and velocity dispersion, $n(\sigma)$, of the VVDS groups are in qualitative agreement with the corresponding $n(z)$ and $n(\sigma)$ recovered from the mock catalogues. The final VVDS group catalogue is characterised by an overall completeness of $\sim60$\% and a purity (fraction of detected groups that actually are a bound structure) of $\sim50$\%.

The group catalogue has been produced using the most secure redshifts (flag = 2, 3, 4, and 9 for primary galaxies, and flag = 22, 23, 24, and 29 for secondary galaxies) without any luminosity or mass selection. In this case we used neither flag = 1 nor 21 galaxies. Although the VVDS selection function described in the next section is well suited to recovering the total number of galaxies in the sample, we used the most accurate redshifts to obtain a reliable group catalogue.

\subsection{Weighting scheme in the VVDS-Deep}\label{complet}
The final VVDS selection function is fully detailed in \citet{vvdsud}, and described in \citet{cucciati12}. Here we only make a short summary.

Since $\sim$25\% of the total number of potential targets in the VVDS-Deep field have been spectroscopically observed and since the redshifts are not measured with 100\% accuracy, we have to correct the number of galaxies in the VVDS-Deep sample using the target sampling rate (TSR) and the spectroscopic success rate (SSR). They have been computed as a function of redshift, source magnitude, and source size ($x$). The SSR is independent of the galaxy type, as demonstrated up to $z \sim 1$ in \citet{zucca06}. Because several first-epoch VVDS-Deep galaxies with flag = 1 and 2 have been observed again in the VVDS-Ultradeep survey ($I_{\rm AB} \leq 24.75$), providing a robust measurement of their redshift, this offers the opportunity to estimate the correct $n(z)$ of VVDS-Deep flag = 1 and 2 sources, and we defined a weight $w_{129}$ to take this into account. We also defined the weight $w_{129}$ for flag = 9 sources by comparison with the latest photometric redshifts in the VVDS-Deep field \citep[see][for details about the latest photometric data set in this field]{cucciati12}. By definition, $w_{129} = 1$ for flag = 3 and 4 sources. We derived the spectroscopic completeness weight for each galaxy $i$ in the catalogue as
\begin{equation}
w^{i}_{\rm spec}(z,I_{\rm AB},x) = \frac{w_{129}^{i}}{TSR^{i} \times SSR^{i}},
\end{equation}
and also defined a weight $w^{k}_{\rm spec} =  w^{\rm prin}_{\rm spec} \times w^{\rm neigh}_{\rm spec}$ for each close neighbour, where $w^{\rm prin}_{\rm spec}$ and $w^{\rm neigh}_{\rm spec}$ are the spectroscopic completeness weights of the principal and the neighbouring galaxy, respectively.

Since the observations were performed under a typical ground-based seeing of $1^{\prime\prime}$, we need to correct for the increasing incompleteness in targeting both components of close pairs because the separation between them is getting smaller. Assuming a clustered distribution of galaxies, the number of galaxy pairs should be a monotonically decreasing function of the pair separation \citep[e.g.,][]{bell06,lin08,deravel09}. However, pairs start to be under-counted for separations $\theta  \leq 2^{\prime\prime}$ because of seeing effects. We apply a weight $w_{\theta}^{k}$ on each close neighbour using the ratio
\begin{equation}
w_{\theta}^{k} = \frac{a}{r_{\rm zz}\,(\theta_k)},
\end{equation}
where $a$ is the probability of randomly selecting a pair, obtained at large separations, and $r_{\rm zz}\,(\theta_k)$ is the ratio between the observed pair count with separation $\theta_k$ in the spectroscopic catalogue ($N_{\rm zz}$) over the observed pair count at the same separation in the photometric one ($N_{\rm pp}$). For large separations ($\theta > 50^{\prime\prime}$), $r_{\rm zz} \sim a$, but at small separations $r_{\rm zz} < a$ because of the artificial decrease in pairs due to seeing effects \citep[see][for further details]{deravel09,clsj11mmvvds}. This weight also accounts for other geometrical biases in the survey, e.g., those related to the minimum separation between spectral slits.

\section{Connecting close neighbours and group environment}\label{richness}

\begin{figure}[t!]
\resizebox{\hsize}{!}{\includegraphics{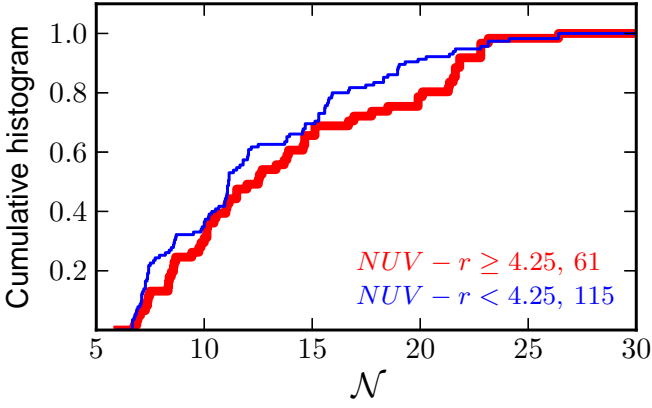}}
\caption{Cumulative histogram of red (thick line) and blue (thin line) principal galaxies as a function of the richness $\mathcal{N}$ of their hosting group. The definition of red and blue galaxy and the number of galaxies in each population are labelled in the figure. [{\it A colour version of this plot is available in the electronic edition}].}
\label{histcolNfig}
\end{figure}

\begin{table}
\caption{Red, passive fraction, $f_{\rm red}$, of principal and companion galaxies as a function of group richness $\mathcal{N}$.}
\label{fredtab}
\begin{center}
\begin{tabular}{lcc}
\hline\hline\noalign{\smallskip}
Richness & $M_B^{\rm e} \leq -20$ & $M_B^{\rm e} \leq -18.25$\\
\noalign{\smallskip}\hline\noalign{\smallskip}
Outside groups	&  $0.21^{+0.02}_{-0.02}$	& $0.12^{+0.01}_{-0.01}$	\\\noalign{\smallskip}
$6 \leq \mathcal{N} < 9$  		&  $0.29^{+0.07}_{-0.06}$ 	& $0.14^{+0.03}_{-0.02}$	\\\noalign{\smallskip}
$9 \leq \mathcal{N} < 15$		&  $0.36^{+0.06}_{-0.05}$ 	& $0.19^{+0.03}_{-0.02}$	\\\noalign{\smallskip}
$15 \leq \mathcal{N} < 30$		&  $0.40^{+0.07}_{-0.06}$ 	& $0.22^{+0.04}_{-0.03}$	\\\noalign{\smallskip}
\hline
\end{tabular}
\end{center}
\end{table}

\subsection{Measuring the group richness}
In this section we describe how we estimated the richness $\mathcal{N}$ of the VVDS-Deep groups and the number of principal galaxies belonging to these groups. We assigned to each group those galaxies in the companion sample that lie within the group's volume $V_{g}$, where $V_{g}$ is defined as the cylinder used in the Phase III of the group finding algorithm. To obtain the total number of galaxies in the group, i.e., their richness $\mathcal{N}$, we summed the weights of the group members:
\begin{equation}
\mathcal{N}^{j} = \sum_{i\ {\rm in}\ V_{g}^{j}} w_{\rm spec}^{i},
\end{equation}
where the index $i$ spans the galaxies in the companion sample\footnote{The principal sample is included in the companion one because of the fainter selection magnitude ($M_{B}^{\rm e} \leq -18.25$) of the companion sample.} and the index $j$ spans the VVDS-Deep groups. If we assigned fewer than two companion galaxies to a given group, we set $\mathcal{N} = 0$. This happened to 114 (30\%) of the groups, leaving 260 groups in VVDS-Deep with $\mathcal{N} \neq 0$ at $0.2 \leq z < 0.95$. In Fig.~\ref{histNfig} we show the redshift distribution of those groups with $\mathcal{N} \geq 5$, $\mathcal{N} \geq 10$ and $\mathcal{N} \geq 15$. The typical DM halo mass of our groups is $M_{\rm h} \sim 10^{13-13.5}\ M_{\odot}$. We estimated this DM halo mass using the relation between the richness of bright galaxies ($M_B^{\rm e} \leq -20$) and $M_{\rm h}$ derived by \citet{knobel09} in the zCOSMOS\footnote{http://www.astro.phys.ethz.ch/zCOSMOS/} \citep{lilly07} survey.

\begin{figure}[t!]
\resizebox{\hsize}{!}{\includegraphics{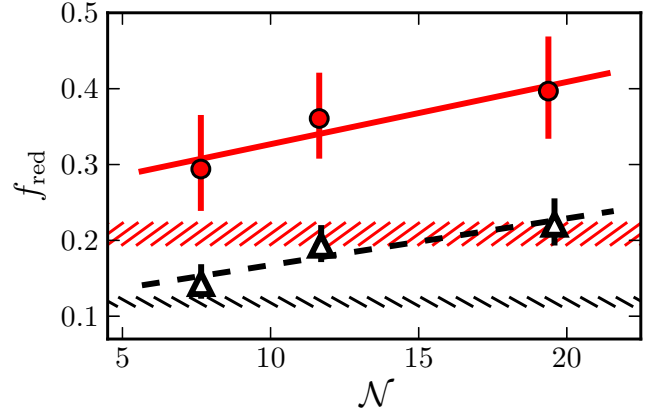}}
\caption{Fraction of red, passive galaxies, $f_{\rm red}$, as a function of group richness $\mathcal{N}$ for principal (dots) and companion (triangles) galaxies. The lines are the best linear least-squares fit to the principal (solid) and companion (dashed) data. The dashed areas show the 68\% confidence intervals of $f_{\rm red}$ for principal (right-dashed) and companion (left-dashed) galaxies outside groups. [{\it A colour version of this plot is available in the electronic edition}].}
\label{fredNfig}
\end{figure}

\begin{figure}[t!]
\resizebox{\hsize}{!}{\includegraphics{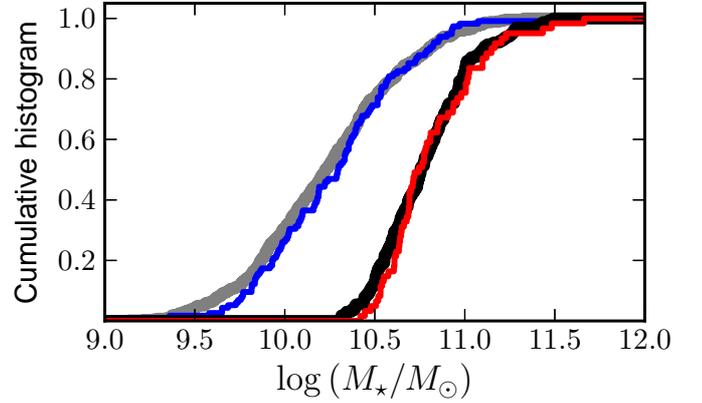}}
\caption{Cumulative histogram of red and blue principal galaxies as a function of the stellar mass. The blue or red (thin) lines are for blue or red principal galaxies hosting by groups, while the the grey or black (thick) lines are for red or blue principals outside groups. [{\it A colour version of this plot is available in the electronic edition}].}
\label{histmassfig}
\end{figure}

We followed the procedure described above to assign the 895 galaxies in the principal sample to each group, both red and blue, 
\begin{equation}
N_{\rm prin}^{j} = \sum_{p\ {\rm in}\ V_{g}^{j}} w_{\rm spec}^{p},
\end{equation}
where the index $p$ spans the red or blue galaxies in the principal sample. We find 197 (22\%) of the principal galaxies in groups, while 176 of those (91\%) are in $\mathcal{N} \neq 0$ groups. In Fig.~\ref{histcolNfig} we show the cumulative histogram of red (61) and blue (115) principal galaxies as a function of the group richness $\mathcal{N}$. The blue principals are more numerous in poor groups, while red principals are mainly in rich ones. This translates to an increase in the red, passive fraction ($f_{\rm red}$) of principal galaxies with $\mathcal{N}$, see Fig.~\ref{fredNfig} and Table~\ref{fredtab}. This has already been shown by \citet{cucciati10gr} with first-epoch VVDS-Deep data. In addition, Fig.~\ref{fredNfig} shows that the red fraction of principal galaxies outside groups is lower than in groups \citep[e.g.,][]{gerke07,balogh09,iovino10,cucciati10gr,presotto12}. These trends are also present in the companion sample, but $f_{\rm red}$ is a factor of two lower than for principal galaxies (Fig.~\ref{fredNfig}). This luminosity dependence is consistent with \citet{iovino10} results.

Finally, we studied the stellar mass of red and blue galaxies. The distribution of the stellar mass of both red and blue principal galaxies hosting by groups and outside groups is similar (Fig.~\ref{histmassfig}). In both colour cases, the mass distribution of principals in groups is described well in logarithmic space by a Gaussian function with median $\mu_{\rm red} = 10.76$ and dispersion $\sigma_{\rm red} = 0.25$ for red galaxies, and median $\mu_{\rm blue} = 10.23$ and dispersion $\sigma_{\rm blue} = 0.43$ for blue galaxies.

\subsection{Assigning close neighbours to groups}
We counted how many neighbours of red or blue principal galaxies belong to each group $j$ with the following equation,

\begin{figure}[t!]
\resizebox{\hsize}{!}{\includegraphics{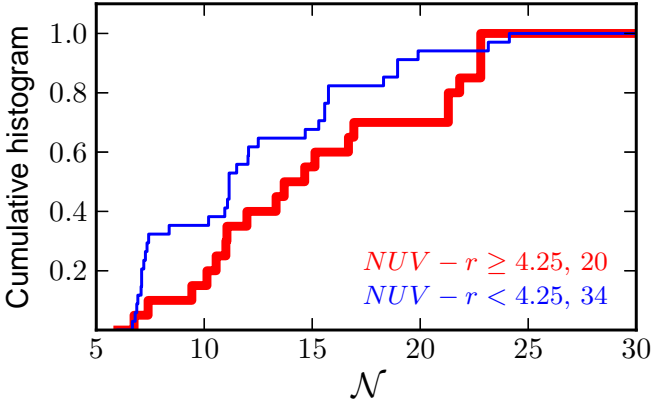}}
\caption{Cumulative histogram of red (thick line) and blue (thin line) principal galaxies with a close neighbour as a function of group richness $\mathcal{N}$. The definition of red and blue galaxy and the number of close neighbours in each population are labelled in the figure. [{\it A colour version of this plot is available in the electronic edition}].}
\label{histffcolNfig}
\end{figure}

\begin{equation}
{N}_{\rm neigh}^{j} = \sum_{k\ {\rm in}\ V_{g}^{j}} w_{\rm spec}^{k} w_{\theta}^{k},
\end{equation}
where the index $k$ spans the close neighbours of the red or blue principals. We find 20 of the 24 neighbours of red galaxies in groups with $\mathcal{N} \neq 0$ (83\%), and 34 of the 40 neighbours of the blue ones (85\%). However, we expected that all our close neighbours belong to groups with at least two members. We checked that in three of the ten orphan neighbours the companion galaxy has flag = 1, which were not used in group determination. The remaining seven neighbours not linked to groups are explained by the performance of the group-finding algorithm, which leads to a completeness in group detection of $\sim60$\% \citep[see][for details]{cucciati10gr}. Taking this into account, the identification of $\sim$90\% of the neighbours in groups supports the reliability of the group-finding algorithm. This detection rate is similar to that in the DEEP2 spectroscopic survey \citep[see][for details]{lin10}. 

We find that blue principals with a close neighbour are mainly located in poor groups, while red principals with a close neighbour appear in the richer ones (Fig.~\ref{histffcolNfig}). This has already been noted by \citet{lin10}. In the next section we study in detail how the number of neighbours per red and blue principal galaxy depends on $\mathcal{N}$.

\begin{figure}[t!]
\resizebox{\hsize}{!}{\includegraphics{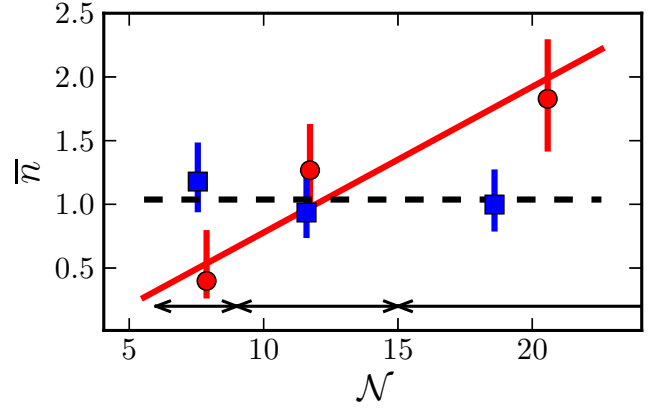}}
\caption{Number of close neighbours per principal galaxy as a function of group richness $\mathcal{N}$. Dots and squares refer to red, passive and blue, star-forming principal galaxies, respectively. The solid line is the best error-weighted least-squares linear fit to the red principal galaxies data. The dashed line is the average number of close neighbours per blue galaxy if it is assumed independent of $\mathcal{N}$. The arrows mark the three ranges of richness probed. [{\it A colour version of this plot is available in the electronic edition}].}
\label{nNfig}
\end{figure}

\begin{table}
\caption{Number of close neighbours ($\overline{n}$) per red, passive and blue, star-forming principal galaxy as a function of group richness $\mathcal{N}$.}
\label{nvsNtab}
\begin{center}
\begin{tabular}{lcccc}
\hline\hline\noalign{\smallskip}
Richness & $\overline{\mathcal{N}}_{\rm red}$ & $\overline{n}_{\rm red}$ & $\overline{\mathcal{N}}_{\rm blue}$ & $\overline{n}_{\rm blue}$\\\noalign{\smallskip}
\hline\noalign{\smallskip}
$6 \leq \mathcal{N} < 9$ 	&  7.9 & $0.40^{+0.40}_{-0.14}$ &  7.5 & $1.18^{+0.31}_{-0.24}$ \\\noalign{\smallskip}
$9 \leq \mathcal{N} < 15$  	& 11.7 & $1.27^{+0.36}_{-0.29}$ & 11.6 & $0.94^{+0.28}_{-0.20}$ \\\noalign{\smallskip}
$15 \leq \mathcal{N} < 30$  	& 20.6 & $1.83^{+0.47}_{-0.41}$ & 18.6 & $1.00^{+0.27}_{-0.21}$ \\\noalign{\smallskip}
\hline\noalign{\smallskip}
\end{tabular}
\end{center}
\end{table}

\section{Number of neighbours per red and blue principal galaxy as a function of $\mathcal{N}$}\label{nvsN}
In this section we estimate the dependence of the number of neighbours on $\mathcal{N}$ and on the colour of the principal galaxy. We define the number of neighbours per principal galaxy in the richness range $\mathcal{N}_{\rm r} = [\mathcal{N}_1, \mathcal{N}_2$) as
\begin{equation}
\overline{n}\,(\mathcal{N}_{\rm r}) = \frac{\sum_{j}^{\mathcal{N}^j \in \mathcal{N}_{\rm r}} N_{\rm neigh}^{j}}{\sum_{j}^{\mathcal{N}^j \in \mathcal{N}_{\rm r}} N_{\rm prin}^{j}}.
\end{equation}
Because each principal galaxy can have more than one close neighbour (see Sect.~\ref{closepairs}) and of the weighting scheme described in Sect.~\ref{complet}, $\overline{n}$ can be higher than one and should not be confused with a (close pair) merger fraction. We estimated the uncertainty in $\overline{n}$ by applying the Bayesian approach described in \citet{cameron11}.

We measured $\overline{n}$ in three richness ranges, $6 \leq \mathcal{N} < 9$, $9 \leq \mathcal{N} < 15$, and $15 \leq \mathcal{N} < 30$. These ranges provide a similar number of red and blue principal galaxies in each bin and ensure good statistics. Because red principal galaxies are primarily found in richer groups (Fig.~\ref{fredNfig}), we estimate the average richness of the groups that host red or blue principal galaxies in the range $\mathcal{N}_{\rm r}$ with the following equation:
\begin{equation}
\overline{\mathcal{N}}\,(\mathcal{N}_{\rm r}) = \frac{\sum_{j}^{\mathcal{N}^j \in \mathcal{N}_{\rm r}} \mathcal{N}^{j} N_{\rm prin}^{j}}{\sum_{j}^{\mathcal{N}^j \in \mathcal{N}_{\rm r}} N_{\rm prin}^{j}}.
\end{equation}

\begin{figure}[t!]
\resizebox{\hsize}{!}{\includegraphics{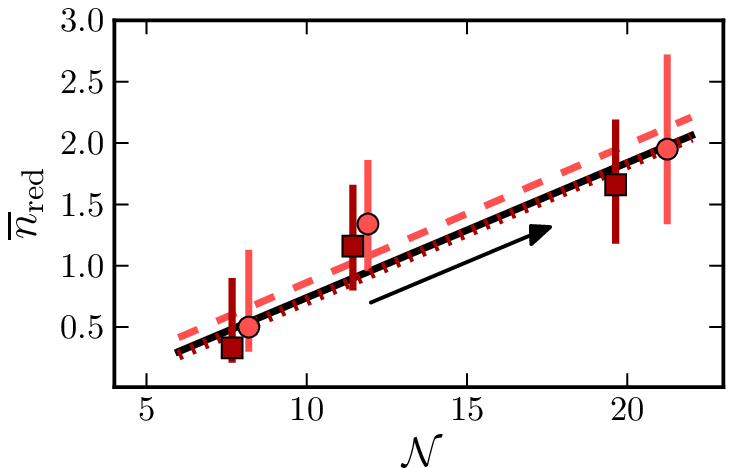}}
\resizebox{\hsize}{!}{\includegraphics{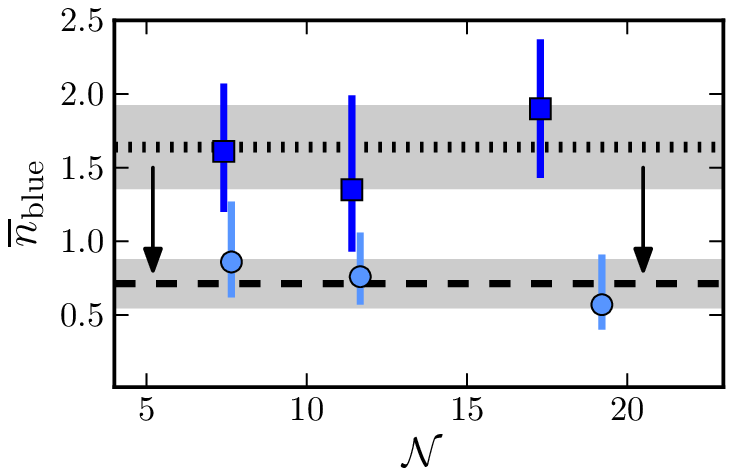}}
\caption{Number of neighbours per principal galaxy as a function of group richness $\mathcal{N}$ and redshift. {\it Top:} $\overline{n}_{\rm red}$ at $0.2 \leq z < 0.65$ (dots) and at $0.65 \leq z < 0.95$ (squares). The dashed and dotted lines are the best error-weighted least-squares linear fit to the data at the lower and higher redshift bins. The solid line is the global relation for red principals in Fig.~\ref{nNfig}, and the arrow marks the redshift evolution of the relation. {\it Bottom:} $\overline{n}_{\rm blue}$ at $0.2 \leq z < 0.81$ (dots) and at $0.81 \leq z < 0.95$ (squares). The dashed and dotted lines are the error-weighted average number of neighbours at the lower and higher redshift bins, with the grey areas showing the uncertainty in these averages. The arrows mark the redshift evolution of the relation. [{\it A colour version of this plot is available in the electronic edition}].}
\label{nNcolfig}
\end{figure}

We summarise our results in Table~\ref{nvsNtab} and show them in Fig.~\ref{nNfig}. We find that
\begin{itemize}
\item The number of neighbours per red, passive galaxy increases with $\mathcal{N}$. The error-weighted least-squares fit of the function $\overline{n}_{\rm red} \propto \alpha_{\rm red}\,\mathcal{N}$ to the data yields $\alpha_{\rm red} = 0.11\pm0.04$. We checked that this positive trend remains significant ($\sim2\sigma$) when different richness bins are used.
\item The number of neighbours per blue, star-forming galaxy is roughly constant with $\mathcal{N}$, $\overline{n}_{\rm blue} = 1.03\pm 0.14$. In this case the slope from the linear fit is consistent with zero, $\alpha_{\rm blue} = -0.01\pm0.03$. As in the previous case, the use of different richness bins does not change this result.
\end{itemize}
That the trends in the number of neighbours for red and blue galaxies are different is the main result of the present paper, and in the next sections we estimate the redshift evolution of these relations.

\begin{table}
\caption{Number of close neighbours per red, passive principal galaxy ($\overline{n}_{\rm red}$) as a function of group richness $\mathcal{N}$ and redshift.}
\label{nvsNredtab}
\begin{center}
\begin{tabular}{lccccc}
\hline\hline\noalign{\smallskip}
Richness & \multicolumn{2}{c}{$\overline{z_{\rm r}}_{,1} = 0.46$} && \multicolumn{2}{c}{$\overline{z_{\rm r}}_{,2} = 0.82$}\\
\cline{2-3}\cline{5-6}\noalign{\smallskip}
                      & $\overline{\mathcal{N}}_{\rm red}$ & $\overline{n}_{\rm red}$ && $\overline{\mathcal{N}}_{\rm red}$ & $\overline{n}_{\rm red}$\\\noalign{\smallskip}
\hline\noalign{\smallskip}
$6 \leq \mathcal{N} < 9$ 	&  8.2 & $0.50^{+0.63}_{-0.20}$ &&  7.7 & $0.33^{+0.57}_{-0.12}$ \\\noalign{\smallskip}
$9 \leq \mathcal{N} < 15$  	& 11.9 & $1.34^{+0.52}_{-0.38}$ && 11.4 & $1.16^{+0.50}_{-0.36}$ \\\noalign{\smallskip}
$15 \leq \mathcal{N} < 30$  	& 21.3 & $1.95^{+0.77}_{-0.61}$ && 19.6 & $1.66^{+0.53}_{-0.48}$ \\\noalign{\smallskip}
\hline\noalign{\smallskip}
\end{tabular}
\end{center}
\end{table}

\begin{table}
\caption{Number of close neighbours per blue, star-forming principal galaxy ($\overline{n}_{\rm blue}$) as a function of group richness $\mathcal{N}$ and redshift.}
\label{nvsNbluetab}
\begin{center}
\begin{tabular}{lccccc}
\hline\hline\noalign{\smallskip}
Richness & \multicolumn{2}{c}{$\overline{z_{\rm r}}_{,3} = 0.60$} && \multicolumn{2}{c}{$\overline{z_{\rm r}}_{,4} = 0.90$}\\
\cline{2-3}\cline{5-6}\noalign{\smallskip}
                      & $\overline{\mathcal{N}}_{\rm blue}$ & $\overline{n}_{\rm blue}$ && $\overline{\mathcal{N}}_{\rm blue}$ & $\overline{n}_{\rm blue}$\\\noalign{\smallskip}
\hline\noalign{\smallskip}
$6 \leq \mathcal{N} < 9$	&  7.6 & $0.86^{+0.41}_{-0.24}$ &&  7.4 & $1.61^{+0.46}_{-0.41}$ \\\noalign{\smallskip}
$9 \leq \mathcal{N} < 15$  	& 11.7 & $0.76^{+0.30}_{-0.19}$ && 11.4 & $1.35^{+0.64}_{-0.42}$ \\\noalign{\smallskip}
$15 \leq \mathcal{N} < 30$  	& 19.2 & $0.57^{+0.34}_{-0.17}$ && 17.3 & $1.90^{+0.47}_{-0.47}$ \\\noalign{\smallskip}
\hline\noalign{\smallskip}
\end{tabular}
\end{center}
\end{table}

\subsection{The redshift evolution of \,$\overline{n}_{\rm red}$ in groups}
To estimate the evolution of the $\overline{n}_{\rm red} - \mathcal{N}$ relation with redshift, we defined two redshift bins that comprise ten neighbours of red principals each, $z_{\rm r,1} = [0.2,0.65)$ and $z_{\rm r,2} = [0.65,0.95)$. We find that the slope $\alpha_{\rm red}$ is constant with redshift, with $\alpha_{\rm red} = 0.11 \pm 0.06$ at $z_{\rm r,1}$ and $\alpha_{\rm red} = 0.11 \pm 0.05$ at $z_{\rm r,2}$ (Table~\ref{nvsNredtab} and Fig.~\ref{nNcolfig}, top panel). In addition, the normalisation of the relation is consistent with a non-evolving value. This fact suggests that red galaxies move along the $\overline{n}_{\rm red} - \mathcal{N}$ relation with cosmic time, i.e., that their number of neighbours increases with the accretion of new members by their hosting group.

A non-evolving $\overline{n}_{\rm red} - \mathcal{N}$ relation is expected by cosmological models. In their work, \cite{mcgee09} study the assembly history of clusters and groups in the semi-analytic model of \citet{font08}. They find that two DM haloes with the same mass but observed at different redshifts have a similar accretion history. The only difference between these two DM haloes is the time scale of the accretion process, which is faster for high redshift clusters and groups (i.e., the available cosmic time to reach a given mass is shorter at high $z$). That is, the accretion histories of two groups with the same richness at different redshifts should be similar, so also a similar number of close neighbours at a given $\mathcal{N}$ is expected for red galaxies irrespective of $z$, as we observe.

In summary, the observed $\overline{n}_{\rm red} \propto 0.11\,\mathcal{N}$ relation does not evolve with redshift, as expected from the semi-analytic model of \citet{font08}. This suggests that red, passive galaxies move along the relation as groups and clusters grow.

\subsection{The redshift evolution of \,$\overline{n}_{\rm blue}$ in groups}\label{nNbluez}
We split our groups into two redshift bins, $z_{\rm r,3} = [0.2,0.81)$ and $z_{\rm r,4} = [0.81,0.95)$. We chose this binning to have a similar number of close neighbours of blue galaxies in both bins. We find that the slope of the $\overline{n}_{\rm blue} - \mathcal{N}$ relation is consistent with zero in both redshift ranges: $\alpha_{\rm blue} = -0.02\pm0.03$ at $\overline{z_{\rm r}}_{,3} = 0.6$ and $\alpha_{\rm blue} = 0.03\pm0.06$ at $\overline{z_{\rm r}}_{,4} = 0.9$. In addition, the average number of neighbours increases with redshift from $\overline{n}_{\rm blue} = 0.71 \pm 0.15$ at $z_{\rm r,3}$ to $\overline{n}_{\rm blue} = 1.64 \pm 0.27$ at $z_{\rm r,4}$ (Table~\ref{nvsNbluetab} and Fig.~\ref{nNcolfig}, bottom panel), a factor of two difference that reflects the fast evolution in the close pair fraction of blue, star-forming galaxies \citep[e.g.,][]{lin08,deravel09,chou10,clsj11mmvvds,clsj13ffmassiv}. We explore the possible physical
explanations of the observed trends and their redshift evolution in Sect.~\ref{discussion}.

\section{The assembly of passive and star-forming $L_B \gtrsim L_B^*$ galaxies in the group environment}\label{discussion}
As we show in Sect.~\ref{intro}, a positive correlation between $\overline{n}$ and $\mathcal{N}$ for a given population implies that this population is located in the DM halo of the group, while a lack of correlation suggests that the population is hosted by DM sub-haloes inside the more massive halo of the group. Based on this fact, our results suggest the picture below for the assembly of passive and star-forming bright galaxies in the group environment.

We find that the number of neighbours of the red, passive principal galaxies, which have a stellar mass $\overline{M_{\star}}_{, {\rm red}}\sim 10^{10.8}\ M_{\odot}$, increases with the richness of the group and that this relation is redshift independent. This seems to indicate that red galaxies ``know'' about the full potential of their hosting group and its assembly history (the larger the number of accreted galaxies, the richer the group and the higher the number of close neighbours of the red galaxies), suggesting that they have been residing in the main DM haloes of the groups for a long time ($z \gtrsim 1$). In their study of the group and cluster assembly in a cosmological contest, \citet{mcgee09} find that a $M_{\rm h} \sim 10^{13.6}\ M_{\odot}$ DM halo (groups with $\mathcal{N} \sim 20$) at $z = 1$ ($z = 0.5$) has more than 50\% of its mass in $M_{\rm h} \geq 10^{13}\ M_{\odot}$ DM haloes (groups with $\mathcal{N} \gtrsim 5$) since $z \sim 1.7$ ($z \sim 1.2$). This is consistent with our picture and supports the idea that massive galaxies were born in or accreted by the group and cluster DM haloes at early times, with a massive red sequence starting building-up in these structures at $z \sim 2$ \citep[i.e.,][]{kodama07,bielby10,strazzullo10,andreon11,tanaka12}. As time goes by, DM haloes hosting the groups are accreting new members from the field, increasing the richness of the groups. Because red galaxies were in place early in the potential well of the groups, their number of neighbours provides a fossil record of matter accretion in the group and leads to the observed correlation between $\overline{n}_{\rm red}$ and $\mathcal{N}$. In this picture, some of these newly accreted members are our blue, star-forming principal galaxies. These blue galaxies have the same number of neighbours irrespective of the group richness, suggesting that this is a recently infalling population from the field. This may also indicate that the parent lower-mass DM haloes in which these blue galaxies reside have become sub-haloes of the larger group structure in which they have fallen.

The time evolution is important in our picture. For example, some time after their infall, blue principal galaxies will finally settle in the group potential, and then might present a positive trend with $\mathcal{N}$ as the red galaxies do; that is, their number of neighbours would increase with the group richness. Instead, a constant number of neighbours with $\mathcal{N}$ is observed at the redshifts we have investigated. This fact, added to the increase on the average $\overline{n}_{\rm blue}$ with $z$, can be explained either by (i) the population of infalling blue galaxies increases with cosmic time, statistically diluting the expected positive trend and the past higher number of companions at lower redshift; or by (ii) the high-redshift blue population disappears because galaxies become red, so only the recently infalling population is visible at low $z$. As the fraction of blue galaxies in groups and clusters decreases with cosmic time, the so-called Butcher-Oemler effect \citep{bo84,depropris03,cucciati10gr,iovino10}, the second scenario seems to be the more likely one: because of some process, the star formation in blue galaxies is quenched, and they become fainter/red, hence no longer observed as the brightest galaxy (our principal sample) but as a companion. In addition, 85\% of our blue principals have $M_{\star} \leq 10^{10.7}\ M_{\odot}$, the mass regime in which environment is expected to notably affect the properties of blue galaxies (Sect.~\ref{intro}). The quenching process could be harassment (the combined effect of multiple high-speed galaxy-galaxy encounters and the interaction with the potential of the group or the cluster, e.g., \citealt{moore96,moore99b}), starvation (the removal of warm and hot gas from the galaxy halo, which cuts off the gas supply, e.g., \citealt{larson80,kawata08}), or ram pressure stripping by the intergroup medium \citep[e.g.,][]{rasmussen06,rasmussen12}. We explore further this scenario in the next section.

We have been studying relatively small groups ($M_{\rm h} \sim 10^{13-13.5}\ M_{\odot}$), so they should accrete even smaller structures from the field, such as pairs or single galaxies. In turn, groups such as those in our sample are accreted by larger groups and clusters. The role of environment in shaping galaxy properties in such small groups, before accretion into clusters, is very important. This is known as pre-processing \citep[see, e.g.,][and references therein]{boselli06}. Some physical processes affecting galaxy properties are more efficient in galaxy groups than in clusters, such as galaxy mergers, thanks to the lower velocity dispersion in groups. The study of galaxy properties within groups is thus strongly related to the understanding of galaxy pre-processing.

\subsection{Infall rate of field galaxies into groups at $z \sim 0.7$}
Motivated by the previous results, and following \citet{peng10}, we develop in this section a simple quenching model in the group environment to estimate the infall rate of field galaxies into groups. We started our model at $\overline{z_{\rm r}}_{,4} = 0.9$ by extracting 1000 random sources $i$ with $M_{B}^{\rm e} \leq -20$ drawn from a Schechter function, obtaining the initial magnitudes $M_{B,i}\,(z = 0.9)$. We used the Schechter function parameters for spectroscopic type 2, 3, and 4 (i.e., star-forming) galaxies in the VVDS-Deep from \citet{zucca06}. We checked that our conclusions are the same if the parameters for blue galaxies in \citet{faber07} are used. Then, we added our sources into the group environment, as an ``infall'', starting the quenching process \citep[e.g.,][]{presotto12}. Quenched galaxies become less luminous in the $B-$band and eventually reach the red sequence after a quenching time $T_{\rm Q}$. We assumed that in this quenching process the blue star-forming galaxy $i$ fades by one magnitude after $T_{\rm Q}$, $M_{B,i}\,(t) = M_{B,i}\,(z = 0.9) + t/T_{{\rm Q},i}$,\footnote{The $B-$band luminosity difference between blue and red galaxies of the same mass in our catalogue is nearly one magnitude, and we assumed and exponential decay in the luminosity with cosmic time, $L_{B}(t) = L_{B}(z = 0.9) \times \exp(-0.4t/T_{\rm Q})$.} where $t$ is the cosmological time since $z = 0.9$. The main hypothesis in our model is the quenching time. As shown by \citet{tinker10}, the quenching time due to environmental processes decreases with redshift as $T_{\rm Q} \propto (1+z)^{-1.5}$, and we took $T_{\rm Q}(z = 0.9) = 1.8 \pm 0.4$ Gyr from their estimations. That is, we assigned a $T_{{\rm Q},i}$ to each random source drawn from a Gaussian with mean 1.8 Gyr and standard deviation 0.4 Gyr. Finally, we estimated at which redshift our galaxies in groups leave the blue principal sample, $z_{\rm sat}$, by becoming either red at $T_{{\rm Q},i}$ (i.e., we assumed an instantaneous transition in the colour) or less luminous than the $M_{\rm B}^{\rm e} \leq -20$ selection.

\begin{figure}[t!]
\begin{center}
\resizebox{\hsize}{!}{\includegraphics{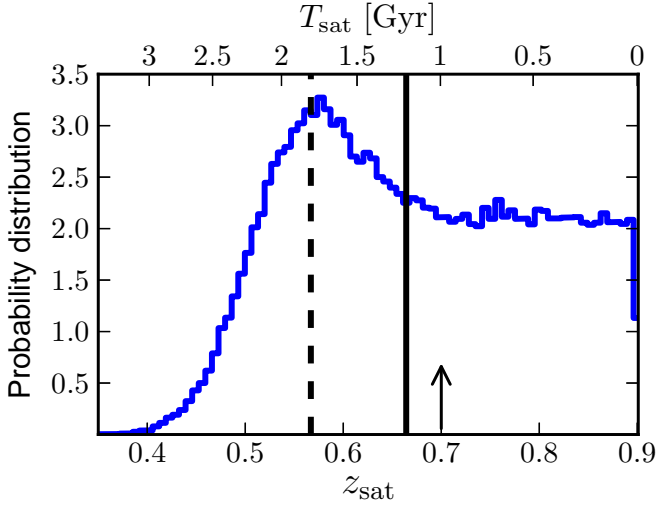}}
\end{center}
\caption{Probability distribution of $z_{\rm sat}$, the redshift at which a blue principal galaxy at $z = 0.9$ becomes a satellite. The upper x-axis shows the cosmological time spanned since $z = 0.9$. The solid vertical line shows the redshift at which half of the initial sources have become a satellite. The dashed line marks $T_{\rm Q} = 1.8$ Gyr after $z = 0.9$. The arrow marks the redshift at which we estimate the infall rate. [{\it A colour version of this plot is available in the electronic edition}].}
\label{zsatfig}
\end{figure}

\begin{figure}[t!]
\begin{center}
\resizebox{\hsize}{!}{\includegraphics{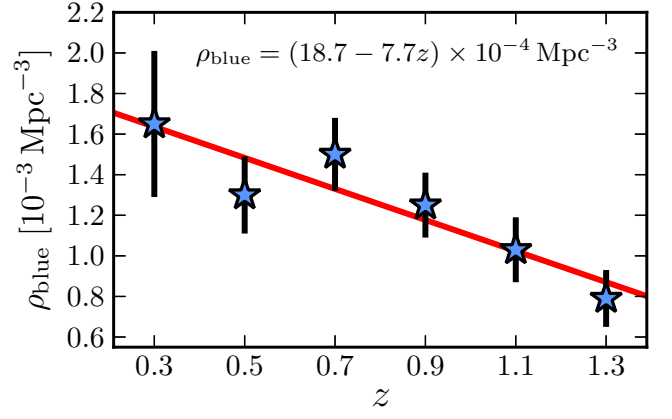}}
\end{center}
\caption{Number density of blue principal galaxies in the VVDS-Deep as a function of redshift. The solid line is the best least-squares linear fit to the data. [{\it A colour version of this plot is available in the electronic edition}].}
\label{nbluefig}
\end{figure}

We show in Fig.~\ref{zsatfig} the probability distribution of $z_{\rm sat}$ from our fiducial model. The probability of leaving the principal sample is nearly constant in the range $z \in (0.6-0.9)$, has its maximum at $z \sim 0.6$, and then starts to decline to $z \sim 0.4$. The maximum is located $T_{\rm Q} = 1.8$ Gyr after the beginning of the model, reflecting the imposed transition to the red sequence. The constant regime is dominated by those galaxies that are still blue but have become too faint to be selected as principals. We estimate that half of the initial sources at $z = 0.9$ have left the principal sample at $z \sim 0.65$, whilst only $\sim 30-35$\% remain at $\overline{z_{\rm r}}_{,3} = 0.6$. Since the number density of blue principals increases from $z = 0.9$ to $z = 0.6$, as we show in Fig.~\ref{nbluefig}, the fading sources must be replaced by new principals infalling from the field. This implies that (i) at $z = 0.6$ the blue principals are dominated by galaxies accreted at $z < 0.9$, in agreement with the observed evolution in the normalisation of the $\overline{n}_{\rm blue} -\mathcal{N}$ relation (Sect.~\ref{nNbluez}), and (ii) an infall rate of $\Re_{\rm infall} \sim 1.1 \times 10^{-4}$ Mpc$^{-3}$ Gyr$^{-1}$ is needed to explain the observed number density of principals at $z = 0.7$. We estimate this infall rate as
\begin{equation}
\frac{\Re_{\rm infall}}{{\rm Mpc}^{-3}\ {\rm Gyr}^{-1}} =  \frac{f_{\rm group}}{T} [\,\rho_{\rm blue}\,(0.7) - (1 - f_{\rm sat})\,\rho_{\rm blue}\,(0.9) ],
\end{equation}
where $\rho_{\rm blue}\,(z)$ is the number density of blue principal galaxies at redshift $z$, $f_{\rm group}$ is the fraction of blue principal galaxies in groups with respect to the total population of blue principals, $T$ is the cosmological time spanned between $z = 0.9$ and $z = 0.7$ in Gyr, and $f_{\rm sat}$ is the fraction of blue principal galaxies at $z = 0.9$ that have become a satellite at $z = 0.7$. We measured $f_{\rm group} = 0.18$ from our data, while our model yields $f_{\rm sat} = 0.41$. To estimate $\rho_{\rm blue}\,(z)$, we performed a linear fit to the number density evolution of blue principal galaxies from \citet{zucca06} in the range $0.2 < z < 1.5$. The obtained fit is $\rho_{\rm blue}\,(z) = (18.7 - 7.7z) \times 10^{-4}\ {\rm Mpc}^{-3}$ (Fig.~\ref{nbluefig}).

The main hypothesis in our toy model is the quenching time, so it is also our main uncertainty. We repeated the previous analysis with $T_{\rm Q} = 1.4$ and 2.2 Gyr, keeping the dispersion of 0.4 Gyr. In the latter case, the peak in $z_{\rm sat}$ distribution is delayed, and $\sim50$\% of the initial sources have left the principal sample by $z = 0.6$. In the former case, only $\sim15$\% of the initial sources survive by $z = 0.6$, and a higher infall rate is needed. From these two extreme models, our best estimation of the infall rate is $\Re_{\rm infall} = 0.9-1.5 \times 10^{-4}$ Mpc$^{-3}$ Gyr$^{-1}$.

In their work, \citet{kovac10} estimate that the fraction of $M_{\star} \sim 10^{10.3}\ M_{\odot}$ galaxies in groups increase with cosmic time with a rate of $f_{\rm infall} \sim7-8$\% Gyr$^{-1}$ because of the infall of field galaxies into the group environment. We also estimate this infall fraction from our model. Assuming that our blue principal galaxies are representative of the $M_{\star} \sim 10^{10.3}\ M_{\odot}$ population, we estimate an infall fraction of $f_{\rm infall} \sim7-10$\% Gyr$^{-1}$, in agreement with \citet{kovac10}. Because of the simplicity of our model, this agreement is remarkable and supports our infall scenario of blue principal galaxies. 

Our toy model has several limitations. For example, infall is a continuous process, so we do not expect peaks in the principal--to--satellite transition, which our toy model produces. In the limiting case of the same $T_{\rm Q}$ for all the galaxies, all the principals reach the red sequence at the same time, and only those that have crossed the selection boundary contribute to the infall rate. Using $T_{\rm Q} = 1.8$ Gyr, we obtain $\Re_{\rm infall} = 1.1 \times 10^{-4}$ Mpc$^{-3}$ Gyr$^{-1}$, similar to our fiducial value. On the other hand, if we increase the dispersion in $T_{\rm Q}$ up to 1 Gyr to spread the peak, we find a probability distribution that decreases with cosmic time. Now several galaxies have low $T_{{\rm Q},i}$ values (even close to zero) and leave the principal sample faster than before. In this case we estimate $\Re_{\rm infall} = 1.3 \times 10^{-4}$ Mpc$^{-3}$ Gyr$^{-1}$, inside our confidence range. Another limitation in our model is that those galaxies that have infalled at $z < 0.9$ also fade and turn red, thus implying a higher infall rate than estimated from our fiducial model. To deal with this limitation, we reduced the redshift range in which we estimated the infall rate. If we use $z = 0.87$ ($\sim 0.15$ Gyr since the beginning of the model) instead of $z = 0.7$ to measure the infall rate, we obtain $\Re_{\rm infall} = 1.3 \times 10^{-4}$ Mpc$^{-3}$ Gyr$^{-1}$, which is consistent with our confidence range. 

In summary, the infall rate from our toy model is mainly determined by the assumed quenching time, and our best estimation is $\Re_{\rm infall} = 0.9-1.5 \times 10^{-4}$ Mpc$^{-3}$ Gyr$^{-1}$. More complex quenching models \citep[i.e.,][]{wetzel13} are needed to fully explore our suggested picture.

\section{Conclusions}\label{conclusion}
We have studied the small-scale environment of $L_B \gtrsim L_B^*$ galaxies hosted by VVDS-Deep groups. We traced the small-scale environment with the number of close neighbours $\overline{n}$, defined as those $M_B^{\rm e} \leq -18.25$ galaxies with a relative distance $5h^{-1}$ kpc $\leq r_{\rm p} \leq 100h^{-1}$ kpc and a relative velocity $\Delta v \leq 500$ km s$^{-1}$ with respect to our principal galaxies ($M_B^{\rm e} \leq -20$). We characterised each group by its richness $\mathcal{N}$, defined as the number of $M_B^{\rm e} \leq -18.25$ galaxies belonging to each group. We split our principal galaxies within $0.2 \leq z < 0.95$ by their rest-frame $NUV - r$ colour into red, passive ($NUV - r \geq 4.25$) and blue, star-forming ($NUV - r < 4.25$).

We find that, on the one hand, the distributions of red and blue principal galaxies in groups as a function of $\mathcal{N}$ are different, showing an increase in the red, passive fraction with $\mathcal{N}$. On the other hand, blue principals with a close neighbour are primarily located in poor groups, while red ones are in rich groups. Combining both results we find that the number of close neighbours per red galaxy increases with $\mathcal{N}$, $\overline{n}_{\rm red} \propto 0.11\,\mathcal{N}$, while that per blue galaxy is roughly constant with $\mathcal{N}$. In addition, these trends are independent of redshift, and only the average $\overline{n}_{\rm blue}$ evolves, decreasing with cosmic time.

Our results support the following assembly history in group environment: red, massive ($M_{\star} \sim 10^{10.8}\ M_{\odot}$) galaxies were either formed or accreted early ($z \gtrsim 1$) in the DM haloes of the groups. Because they are already settled in the group potential at $z \lesssim 1$, the number of neighbours of red, passive galaxies is a record of the accretion history of the group, traced by its richness $\mathcal{N}$. On the other hand, blue, less massive ($M_{\star} \sim 10^{10.3}\ M_{\odot}$) galaxies are an infalling population from the field. They have been accreted by the group potential recently and are still located in their parent DM sub-halo, having the same number of neighbours irrespective of $\mathcal{N}$. As time goes by, these blue star-forming galaxies settle in the group potential and turn fainter and/or red, thus becoming satellite galaxies. Thanks to a simple quenching model, we estimate an infall rate of field galaxies into the group environment of $\Re_{\rm infall} = 0.9-1.5 \times 10^{-4}$ Mpc$^{-3}$ Gyr$^{-1}$ at $z \sim 0.7$.

This proposed picture makes some specific predictions that have to be explored in the future. First, the $\overline{n} - \mathcal{N}$ relation should be present, for both red and blue galaxies, in low-redshift groups. Second, a constant $\overline{n}_{\rm blue}$ as a function of the normalised distance to the centre of the group might be expected, whilst an increase in $\overline{n}_{\rm red}$ towards the group potential centre should be observed. This could be tested using the lower redshift GAMA\footnote{http://www.gama-survey.org/} (Galaxy and Mass Assembly, \citealt{driver11}; $z \sim 0.1$) survey or the low-redshift end of zCOSMOS ($0.2 \leq z \leq 0.5$). Finally, the study of the $\overline{n} - \mathcal{N}$ relation in mass-selected samples is needed to better understand the quenching mechanism that operates in the group environment.

\begin{acknowledgements}
We dedicate this paper to the memory of our six IAC colleagues and friends who
met with a fatal accident in Piedra de los Cochinos, Tenerife, in February 2007,
with special thanks to Maurizio Panniello, whose teachings of \texttt{python}
were so important for this paper. We thank the anonymous referee for the comments
and suggestions.\\

This work is supported by funding from ANR-07-BLAN-0228 and ERC-2010-AdG-268107-EARLY.

\end{acknowledgements}

\bibliography{biblio}
\bibliographystyle{aa}

\end{document}